\pdfoutput=1 
\documentclass[letterpaper,english,aps,prl,floatfix,reprint,superscriptaddress,showpacs,twocolumn]{revtex4-2}
\usepackage[T1]{fontenc}
\usepackage[utf8]{inputenc}
\usepackage{amsmath}
\usepackage{amssymb}
\usepackage{graphicx}

\makeatletter

\usepackage{mathrsfs}
\usepackage{bm}
\usepackage{xspace}
\allowdisplaybreaks[4]  

\usepackage[colorlinks,linkcolor=blue,anchorcolor=blue,citecolor=blue]{hyperref}
\usepackage{ragged2e}
\makeatletter
\AtBeginDocument{%
  \long\def\@makecaption#1#2{%
    \par
    \vskip\abovecaptionskip
    \begingroup
      \small\normalfont
      \setlength{\parindent}{0pt}%
      \setlength{\leftskip}{0pt}%
      \setlength{\rightskip}{0pt}%
      \setlength{\parfillskip}{0pt plus 1fil}%
      \emergencystretch=2em
      \noindent\justifying
      \@make@capt@title{#1}{#2}\par
    \endgroup
    \vskip\belowcaptionskip
  }%
  \def\@caption@fignum@sep{: \ }%
}
\makeatother

\usepackage[caption=false]{subfig}

\usepackage{babel}
\begin{document}
\title{Eigenvalue Statistics of Random Quantum Geometry}
\author{Mingpu Jiang}
\affiliation{Anhui Provincial Key Laboratory of Low-Energy Quantum Materials and
Devices, High Magnetic Field Laboratory, HFIPS, Chinese Academy of
Sciences, Hefei, Anhui 230031, China}
\affiliation{Department of Physics, University of Science and Technology of China,
Hefei 230026, P.R. China}
\author{Jianhui Zhou}
\email{jhzhou@hmfl.ac.cn}
\affiliation{Anhui Provincial Key Laboratory of Low-Energy Quantum Materials and
Devices, High Magnetic Field Laboratory, HFIPS, Chinese Academy of
Sciences, Hefei, Anhui 230031, China}
\begin{abstract}
The quantum geometric tensor is a fundamental property of quantum
states, with broad applications in condensed matter physics, topological
phases, and quantum phase transitions. The eigenvalues characterize
the scale, anisotropy, and effective rank of random quantum geometry,
going beyond scalar quantities such as the trace. Here we study the
eigenvalue statistics of the quantum geometric tensor in finite-dimensional
parameter-dependent random Hamiltonians. We obtain exact analytical
results for the first two nontrivial cases, $N=2$ and $N=3$, with
$N=3$ already showing genuine shape fluctuations. We further propose
a finite-$N$, arbitrary-$D$ description of QGT eigenvalue statistics
and verify it by numerical simulations. Our results provide exact
benchmarks and a practical framework for random quantum geometry in
finite-dimensional disordered and chaotic systems.
\end{abstract}
\maketitle

\section{introduction}
The quantum geometric tensor (QGT), which unifies the Fubini-Study
metric and the Berry curvature \citep{provost1980riemannian,Xiao2010berry,wilczek1989geometric,bengtsson2017geometry},
is a central geometric object in condensed matter physics, topological
phases, and quantum phase transitions \citep{gianfrate2020measurement,torma2023essay,zanardi2007information,piechon2016geometric}.
In complex quantum systems, such as disordered or chaotic systems,
quantum geometry is generally not a fixed deterministic tensor but
a fluctuating object tied to the statistics of eigenstates. Parameter-dependent
random Hamiltonians provide a minimal and universal setting for describing
these fluctuations, because they retain the perturbative structure
of the QGT while replacing microscopic details by random matrix statistics
\citep{niu1985quantized,dyson1962brownian,bohigas1984characterization,guhr1998random,andreanov2025dyson,beenakker1997random,filippone2016drude,werner2019universal,monthus2017many,leblond2021universality,romeral2025scaling}.

Previous works have shown that QGT-related observables in random Hamiltonians
exhibit nontrivial statistical behavior \citep{Walker1995universal,berry2020quantum,berry2020geometric,gat2021correlations,sierant2019fidelity,penner2021hilbert}.
However, most existing results focus on scalar quantities, such as
the trace of the QGT, or on low-dimensional parameter spaces \citep{berry2020quantum,berry2020geometric,sierant2019fidelity}.
A systematic arbitrary-$D$ theory for the eigenvalue statistics of
the QGT remains absent.

The eigenvalues of the QGT are the natural variables for characterizing
random quantum geometry. They encode the total scale, anisotropy,
principal sensitivities, and effective rank of the geometric response,
whereas scalar observables such as the trace only measure the overall
magnitude \citep{medvidovic2023variational,reilly2023optimal}. This
distinction becomes especially important in high-dimensional parameter
spaces, where the same trace may correspond to either an almost isotropic
geometry or a strongly anisotropic one dominated by a few directions
\citep{park2020geometry,machta2013parameter}.

In this work, we develop an arbitrary-$D$ theory for the eigenvalue
statistics of the QGT in parameter-dependent random Hamiltonian families.
We show that the QGT distribution is invariant under unitary rotations
in parameter space, which reduces the problem to the joint statistics
of its eigenvalues. We obtain exact analytical results for the first
two nontrivial cases, $N=2$ and $N=3$. The $N=2$ case gives a rank-one
QGT and serves as the elementary benchmark, while the $N=3$ case
is the first case with genuine shape fluctuations. Guided by these
exact solutions, we further construct a finite-$N$, arbitrary-$D$
model for the QGT eigenvalue distribution. Numerical simulations agree
well with the analytical results and support the general description.

The remainder of this paper is organized as follows. In Sec. II we
introduce the parameter-dependent random-Hamiltonian model and establish
the unitary invariance of the QGT distribution in parameter space.
Sec. III and IV present the exact eigenvalue statistics for the first
two nontrivial cases, $N=2$ and $N=3$, respectively. In Sec. V we
use these exact results to construct a finite-$N$, arbitrary-$D$
model for the QGT eigenvalue distribution. Sec. VI discusses a resonant-disordered
realization of the finite-dimensional mixed-GUE model, showing how
such random Hamiltonians can emerge as effective descriptions of rare
resonant blocks in strongly disordered systems. Sec. VII concludes
with a discussion of possible extensions.

\section{model and spectral reduction}
Consider the parameter-dependent random Hamiltonian family
\begin{align}
H & =H_{0}+\sum_{i=1}^{D}y_{i}H_{i}\\
p(H_{i}) & \propto\mathrm{Exp}(-\frac{N}{2}\mathrm{tr}H_{i}^{2}).
\end{align}
where $\{H_{0},H_{i}\}$ are independent and identically distributed
$N\times N$ Gaussian unitary ensemble (GUE) matrices. When $D\leq3$,
this ensemble reproduces to the low-dimensional random-Hamiltonian
models studied previously in \citep{berry2020quantum,berry2020geometric,gat2021correlations,sierant2019fidelity,penner2021hilbert}.
At the origin of parameter space, the QGT associated with a non-degenerate
eigenstate $|n\rangle$ of $H_{0}$ is
\begin{equation}
G_{\alpha\beta}^{(n)}=\sum_{m=1,\not=n}^{N}\frac{\langle n|H_{\alpha}|m\rangle\langle m|H_{\beta}|n\rangle}{(E_{n}-E_{m})^{2}}
\end{equation}
with $\alpha,\beta\in\{y_{i},i=1,...,D\}$. This representation separates
the random transition matrix elements in the numerator from the level-spacing
factors in the denominator, and it forms the starting point for the
analysis below.

One can show that the distribution of $G^{(n)}$ is invariant under
unitary rotations in parameter space. Indeed, we define a set of $D$-dimensional
complex vectors $\{X_{m}\},(X_{m})_{\alpha}\equiv\frac{(H_{\alpha})_{nm}}{E_{m}-E_{n}}.$
Using Eq. (2), the QGT can be written as
\begin{equation}
G^{(n)}=\sum_{m=1,\not=n}^{N}X_{m}(X_{m})^{\dagger}.
\end{equation}

When $H_{0}$ is fix, $\{X_{m}\}$ are independent centered complex
Gaussian vectors with conditional covariance
\begin{equation}
\mathbb{E}[(X_{m})_{\alpha}(X_{m'})_{\beta}^{*}|H_{0}]=\frac{\delta_{mm'}\delta_{\alpha\beta}}{N(E_{m}-E_{n})^{2}}.
\end{equation}

The conditional covariance is isotropic in parameter space, thus the
joint Gaussian density is invariant under $U(D)$ rotations. $(UX_{m})\stackrel{d}{=}X_{m}$,
where $\stackrel{d}{=}$ denotes equality in distribution. Hence,
conditioned on $H_{0}$, the distribution of QGT is $U(D)$-invariance:
\begin{equation}
UG^{(n)}U^{\dagger}\stackrel{d}{=}G^{(n)},\mathrm{conditioned}\ \mathrm{on}\ H_{0}.
\end{equation}

Because the numerators are generated by perturbation matrices independent
of $H_{0}$, whereas the denominators depend only on the level spacings
of $H_{0}$, averaging over $H_{0}$ randomizes only the radial weights
$(E_{m}-E_{n})^{-2}$ and does not break the isotropy in parameter
space. Hence the conditional $U(D)$-invariance immediately extends
to the full distribution.
\begin{equation}
UG^{(n)}U^{\dagger}\stackrel{d}{=}G^{(n)}.
\end{equation}

It states that, at the ensemble level, the orientation of the QGT
in parameter space has no preferred direction. The nontrivial, rotation-invariant
information is therefore contained in the nonzero eigenvalues of $G^{(n)}$.
In this respect, the statistics of random QGTs reduce naturally to
a spectral problem.

The spectral reduction also motivates a scale-shape parametrization
of the nonzero QGT eigenvalues. Let $\lambda_{1},...,\lambda_{r}$
be the nonzero eigenvalues of the QGT, with $r\equiv\mathrm{rank}(G)=\mathrm{min}(N-1,D)$.
We define the scale variable and the normalized shape variables by
$S\equiv\sum_{i=1}^{r}\lambda_{i},\ \xi_{i}\equiv\frac{\lambda_{i}}{\sum_{i=1}^{r}\lambda_{i}}$.
Here $S$ measures the overall strength of the QGT, while $\xi_{i}$
describe how this strength is distributed among the principal directions
in parameter space. Including the Jacobian of the transformation from
$\lambda_{i}$ to $(S,\xi_{i})$, the joint eigenvalue density can
be written as
\begin{equation}
p_{G}(\lambda_{1},...,\lambda_{r})=\frac{1}{s^{r-1}}p_{S}(s)p_{\xi}(\xi|s).
\end{equation}

In the following sections, this decomposition will be used as the
common language for all finite-$N$ results. The two-level case $N=2$
provides the basic benchmark: the QGT has rank one, so the shape sector
is frozen and only the scale distribution remains nontrivial. The
first genuinely nontrivial shape distribution appears at $N=3$. A more systematic derivation is given in Appendix A.

\section{exact eigenvalue distributions for $N=2$ }
\begin{figure}
\includegraphics[width=\columnwidth]{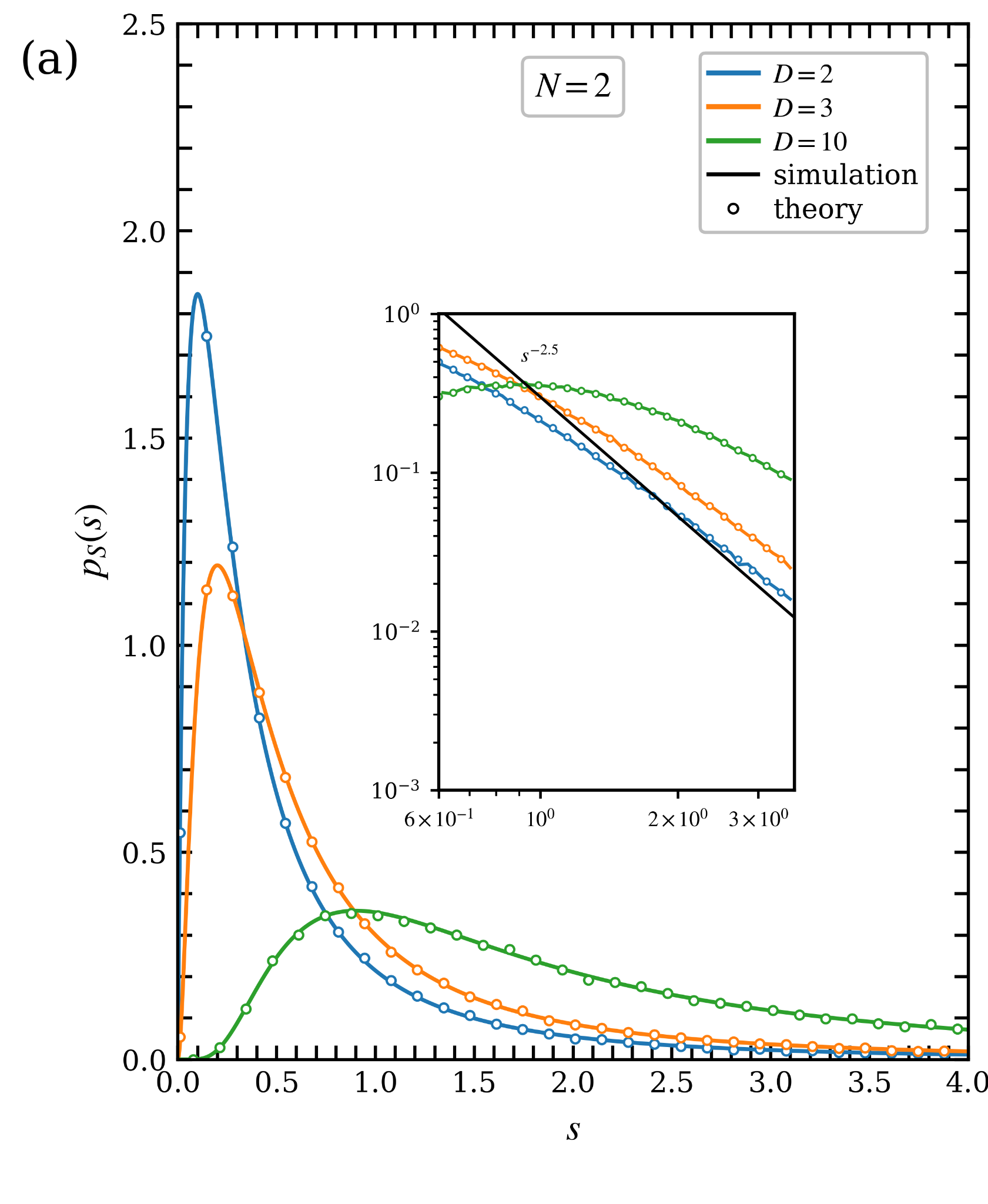}
\begin{turnpage}
\caption{\label{fig:n2} Exact scale distribution for the two-level random-Hamiltonian
model. The density $p_{S}(s)$ of the single nonzero QGT eigenvalue
is shown for $\ensuremath{D=2,3,10}.$ Solid curves denote Monte Carlo
simulations, and open symbols denote the exact result in Equation
(12). The inset shows the universal large-$S$ tail $p_{S}(s)\sim s^{-5/2}$.}
\end{turnpage}
\end{figure}

The two-level problem provides the minimal finite-$N$ benchmark for
QGT eigenvalue statistics. Since there is only one virtual transition
channel, the QGT has rank one for arbitrary $D$. Thus the shape sector
is trivial, whereas the scale distribution is already nontrivial and
contains the basic head-tail structure that reappears at higher $N$.
For $N=2$, $r=1,\,S=\lambda_{max}.$

We consider the QGT associated with the first eigenstate,
\begin{equation}
G_{\alpha\beta}^{(1)}=\frac{\langle1|H_{\alpha}|2\rangle\langle2|H_{\beta}|1\rangle}{(E_{1}-E_{2})^{2}}.
\end{equation}

Defining $\omega_{\alpha}\equiv\langle1|H_{\alpha}|2\rangle,\,\omega=(\omega_{1},..,\omega_{D})^{T}$,
we can rewrite $G^{(1)}$ as $G^{(1)}=\frac{\omega\omega^{\dagger}}{(E_{1}-E_{2})^{2}}.$
Since $r=1$, there is single nonzero eigenvalue, denoted by $\lambda_{max}$
\begin{align}
\lambda_{max} & =\frac{1}{(E_{1}-E_{2})^{2}}(\omega^{\dagger}\omega)\nonumber \\
 & =\frac{1}{(E_{1}-E_{2})^{2}}(\sum_{\alpha=1}^{D}|\langle1|H_{\alpha}|2\rangle|^{2}).
\end{align}

The numerator is the total transition strength generated by all parameter
directions, while the denominator is the squared level spacing. Large
QGT events are therefore associated with near-degenerate levels, whereas
small QGT events require all transition amplitudes to be simultaneously
suppressed.

The shape sector $\xi$ is degenerate, 
\begin{equation}
p_{\xi}(\xi|s)=\delta_{\xi,1},
\end{equation}
where $\xi$ is independent of $S$ when $N=2$.

To obtain $p_{S}(s)$, it is convenient to define $K\equiv\sum_{\alpha=1}^{D}|\langle1|H_{\alpha}|2\rangle|^{2},\,U\equiv\Delta^{2}\equiv(E_{2}-E_{1})^{2}$,
Then $\lambda_{max}\equiv\frac{K}{U}$,$\,K\stackrel{d}{=}\Gamma(D,2),\,U\stackrel{d}{=}\Gamma(\frac{3}{2},\frac{1}{2})$
are independent Gamma random variables. In Appendix B, we derive the
probability density of $S$, getting the beta-prime distribution
\begin{equation}
p_{S}(s)=\frac{4^{D}}{B(D,\frac{3}{2})}s^{D-1}(1+4s)^{-(D+\frac{3}{2})},
\end{equation}
where $B$ is the beta function. 

Equation (12) gives the complete scale statistics of the rank-one
QGT in the two-level problem. We test this result in Fig.~\ref{fig:n2} by
comparing the exact density with Monte Carlo simulations for several
parameter dimensions, $\ensuremath{D=2,3,10}.$ The agreement is essentially
exact over the full plotted range. Increasing $D$ shifts the bulk
of the distribution to larger $S$, because the numerator $K=\sum_{\alpha=1}^{D}|\langle1|H_{\alpha}|2\rangle|^{2}$
collects transition strength from more parameter directions. By contrast,
the large-$S$ tail is independent of $D$ and follows $p_{S}(s)\sim s^{-5/2},$
as shown in the inset. This confirms that large QGT events are controlled
by the universal small-spacing behavior of the two-level GUE spectrum,
while the parameter dimension mainly controls the small-$S$ head
and the bulk scale.

The $N=2$ solution therefore serves as the first exact benchmark
for random quantum geometry. It cleanly separates scale from shape,
shows that the shape sector is frozen in the rank-one case, and identifies
beta-prime statistics as the natural analytic structure governing
the scale fluctuations. However, because only one nonzero eigenvalue
is present, the $N=2$ problem cannot describe anisotropy or shape
fluctuations. The first case with a genuinely fluctuating QGT eigenvalue
shape is $N=3$, to which we now turn.

\section{exact eigenvalue distributions for $N=3$}
\begin{figure}
\includegraphics[width=\columnwidth]{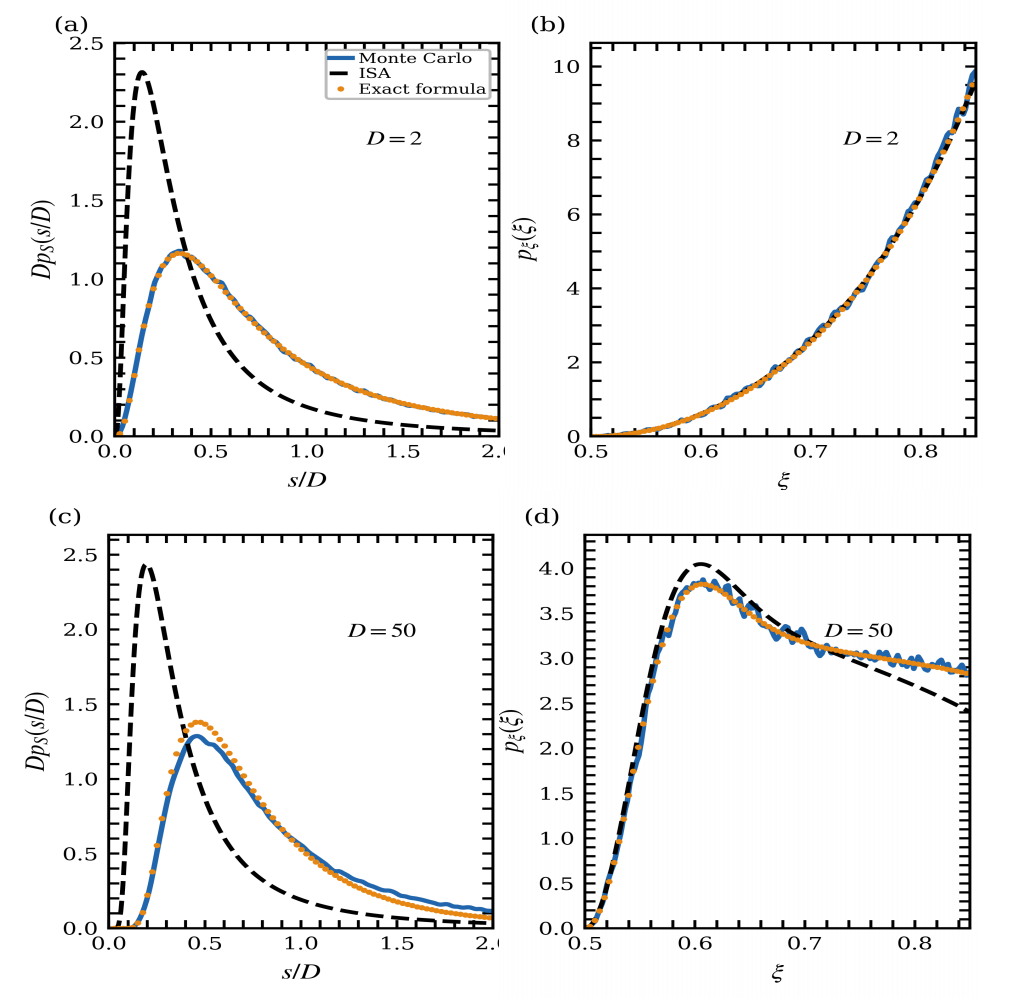}
\begin{turnpage}
\caption{\label{fig:n3} Scale and shape statistics of the QGT for $N=3$.
Panels $(a,c)$ show the rescaled trace distribution $Dp_{S}(s/D)$,
and panels $(b,d)$ show the shape distribution $p_{\xi}(\xi)$, for
$D=2$ and $D=50$. Monte Carlo simulations are compared with the
exact finite-$N$ theory and the independent-spacing approximation
(ISA).}
\end{turnpage}
\end{figure}

\subsection{The density of eigenvalues of QGT conditioned on $H_{0}$}
For $N=3$, the QGT of the middle eigenstate receives contributions
from two neighboring levels. The distribution therefore contains a
nontrivial shape sector, which describes the relative weight carried
by another two principal directions of the QGT.

We focus on the middle state $|2\rangle$ and define $\omega_{\alpha}^{(i)}\equiv\langle2|H_{\alpha}|i\rangle,\ \omega^{(i)}\equiv(\omega_{1}^{(i)},..,\omega_{D}^{(i)})^{T},\ \Delta_{i}\equiv|E_{2}-E_{i}|,\ i=1,3$.
The QGT is the sum of two correlative channels
\begin{equation}
G_{\alpha\beta}^{(2)}\equiv\frac{\omega_{\alpha}^{(1)}\omega_{\beta}^{(1)\dagger}}{(\Delta_{1})^{2}}+\frac{\omega_{\alpha}^{(3)}\omega_{\beta}^{(3)\dagger}}{(\Delta_{3})^{2}}.
\end{equation}

To separate amplitude and orientation, we introduce
\[
X^{(1)}\equiv\frac{|\omega^{(1)}|^{2}}{\Delta_{1}^{2}},\,X^{(3)}\equiv\frac{|\omega^{(3)}|^{2}}{\Delta_{3}^{2}},
\]
\begin{equation}
u^{(1)}\equiv\frac{\omega^{(1)}}{|\omega^{(1)}|},\,u^{(3)}\equiv\frac{\omega^{(3)}}{|\omega^{(3)}|},\,\phi\equiv|u^{(1)\dagger}u^{(3)}|^{2}.
\end{equation}

Here $X^{(1)},\,X^{(3)}$ measure the strength of two fluctuation
channels, while $\phi$ measures their relative orientation in parameter
space. In Appendix C, we derive the for the eigenvalues of the QGT
explicitly. The two nonzero eigenvalues of the QGT are 
\begin{equation}
\lambda_{\pm}=\frac{1}{2}(X^{(1)}+X^{(3)}\pm\sqrt{(X^{(1)}-X^{(3)})^{2}+4X^{(1)}X^{(3)}\phi}).
\end{equation}

As in previous section, we introduce scale and shape variables, $S\equiv\lambda_{+}+\lambda_{-},\,\xi\equiv\frac{\lambda_{+}}{S}$.
The shape fluctuation in the $N=3$ problem has a clear physical origin:
it arises from the competition between two virtual transition channels
and their relative orientation in parameter space.

To derive the exact distribution, it is convenient to proceed in two
steps. We first fix the two adjacent level spacings $\Delta_{1}$
and $\Delta_{3}$, and derive the conditional distribution $p_{S,\xi}(s,\xi|\Delta_{1},\Delta_{3})$.
At fixed gaps, the fluctuation strengths and the angular variable
simplify considerably: $X^{(1)}$ and $X^{(3)}$ become independent
Gamma variables, while the orientation variable $\phi$ follows a
simple Beta distribution. The remaining coupling is the mixing between
scale and shape can be handled by introducing the relative channel
weight $z=\frac{X^{(1)}}{X^{(1)}+X^{(3)}}$. At fixed $\Delta_{1},\Delta_{3}$,
the density factorizes into a scale part and an angular part in the
auxiliary variable:
\begin{align}
p_{S,z}(s,z|\Delta_{1},\Delta_{3}) & =(\Delta_{1}\Delta_{3})^{2D}\frac{3^{2D}}{\Gamma^{2}(D)}(z(1-z))^{D-1}s^{2D-1}\nonumber \\
 & \times\mathrm{Exp}(-3(\Delta_{1}^{2}sz+\Delta_{3}^{2}s(1-z)))\nonumber \\
p_{\xi}(\xi,z) & =(D-1)[\xi(1-\xi)]^{D-2}(2\xi-1)\nonumber \\
 & \times[\frac{1}{z(1-z)}]^{D-1}.
\end{align}

Integrating over $z$ gives
\begin{align}
 & \phantom{}p_{S,\xi}(s,\xi|\Delta_{1},\Delta_{3})\nonumber \\
 & =\frac{(D-1)(\Delta_{1}\Delta_{3})^{2D}3^{2D-1}s^{2D-2}}{\Gamma(D)^{2}(\Delta_{1}^{2}-\Delta_{3}^{2})}\nonumber \\
 & \phantom{}\times[\xi(1-\xi)]^{D-2}(2\xi-1)[\mathrm{Exp}(-3(\Delta_{1}^{2}s+(\Delta_{3}^{2}-\Delta_{1}^{2})s\xi))\nonumber \\
 & \phantom{}-\mathrm{Exp}(-3(\Delta_{1}^{2}s\xi+\Delta_{3}^{2}s(1-\xi))))].
\end{align}

The derivation of this expression is given in Appendix C. It already
shows the essential difference from $N=2$: the scale and shape sectors
are no longer trivially separated, because both are influenced by
the two neighboring gaps.

\subsection{Averaging over energy spacings}
The second stage is to average over the gap statistics of the $3\times3$
GUE matrix itself. Writing the three eigenvalues in terms of their
center and the two adjacent gaps, one finds the joint distribution
\begin{align}
p_{\Delta}(\Delta_{1},\Delta_{3}) & =\frac{81}{4\sqrt{3}\pi}\Delta_{1}^{2}\Delta_{3}^{2}(\Delta_{1}+\Delta_{3})^{2}\nonumber \\
 & \phantom{}\times\mathrm{Exp}(-(\Delta_{1}^{2}+\Delta_{3}^{2}+\Delta_{1}\Delta_{3})).
\end{align}

Substituting the conditional density into the gap average and changing
variables to $x\equiv\Delta_{1}+\Delta_{3}$ and $y\equiv(\Delta_{1}-\Delta_{3})/(\Delta_{1}+\Delta_{3})\in[-1,1]$,
we obtain
\begin{align}
 & \phantom{}p_{S,\xi}(s,\xi)\nonumber \\
 & =K_{D}s^{2D-2}[\xi(1-\xi)]^{D-2}(2\xi-1)\nonumber \\
 & \phantom{}\times\int_{0}^{\infty}dx\int_{-1}^{1}dyx^{4D+5}(1-y^{2})^{2D+2}\nonumber \\
 & \times\mathrm{Exp}(-\frac{x^{2}}{4}[3(1+s)+(1+3s)y^{2}])\frac{\mathrm{sinh}(\frac{3}{2}sx^{2}y(2\xi-1))}{y},
\end{align}
where $K_{D}=\frac{(D-1)3^{2D+3}}{2^{4D+6}\sqrt{3}\pi\Gamma(D)^{2}}.$
Averaging over the transition amplitudes and the correlated adjacent
spacings of the $3\times3$ GUE spectrum gives the exact joint distribution
$p_{S,\xi}$. Using Euler's integral formula \citep{abramowitz1948handbook},
the final result can be written as a convergent hypergeometric series:
\[
p_{S,\xi}(s,\xi)=[\xi(1-\xi)]^{D-2}\sum_{l=0}^{\infty}c_{l}(s)(2\xi-1)^{2l+2},
\]
where
\begin{align}
c_{l}(s) & =\frac{(D-1)3^{2D+2l+4}2^{2l}\Gamma(2D+2l+4)}{\sqrt{3}\pi\Gamma(D)^{2}(2l+1)!}\nonumber \\
 & \phantom{}\times s^{2D+2l-1}[3(1+s)]^{-(2D+2l+4)}B(l+\frac{1}{2},2D+3)\nonumber \\
 & \phantom{}\times{}_{2}F_{1}(2D+2l+4,l+\frac{1}{2};l+2D+\frac{7}{2},-\frac{1+3s}{3(1+s)}),
\end{align}
and $_{2}F_{1}$ is the hypergeometric function.

\subsection{Simple gap approximations}
The exact $N=3$ formula not only acts as a benchmark, but also as
a guide for constructing controlled approximations. The main technical
difficulty lies in the joint distribution of the two adjacent gaps
$\Delta_{1}$ and $\Delta_{3}$. We therefore introduce simplified
gap sectors to isolate the contributions of two-channel geometry and
adjacent-gap correlations to the QGT statistics. Two approximations
are particularly instructive.

In the equal-spacing approximation (ESA), we set the two adjacent
gaps equal, $\Delta_{1}=\Delta_{3}$. Physically, this approximation
suppresses fluctuations of the gap ratio and retains only the symmetric
two-channel geometry. The corresponding gap distribution is taken
as 
\begin{equation}
p_{\Delta}^{ESA}(\Delta_{1},\Delta_{3})\equiv\sqrt{\frac{2}{\pi}}\Delta_{1}^{2}\mathrm{Exp}(-\frac{\Delta_{1}^{2}}{2})\delta_{\Delta_{1},\Delta_{3}},
\end{equation}
which corresponds to 
\begin{align}
p_{S,\xi}^{ESA}(s,\xi) & =\frac{6^{2D}\Gamma(2D)}{B(2D,\frac{3}{2})\Gamma(D)\Gamma(D-1)}s^{2D-1}(1+6s)^{-(2D+\frac{3}{2})}\nonumber \\
 & \phantom{}\times[\xi(1-\xi)]^{D-2}(2\xi-1)^{2}.
\end{align}

The independent-spacing approximation (ISA) was introduced to compute
the trace distribution \citep{berry2020quantum}. Here we extend this
idea to the eigenvalue distribution. In contrast to the ESA, one keeps
the two gap variables but neglects their correlations, replacing the
exact joint density by a product form,
\begin{equation}
p_{\Delta}^{ISA}(\Delta_{1},\Delta_{3})=\frac{2}{\pi}\Delta_{1}^{2}\Delta_{3}^{2}\mathrm{Exp}(-\frac{\Delta_{1}^{2}+\Delta_{3}^{2}}{2}),
\end{equation}
which gives
\begin{align}
p_{S,\xi}^{ISA}(s,\xi) & =\frac{(D-1)3^{2D}\Gamma^{2}(D+\frac{3}{2})}{2\pi\Gamma^{2}(D)}s^{2D-1}(\frac{1+3s}{2})^{-(2D+3)}\nonumber \\
 & \phantom{}\times[\xi(1-\xi)]^{D-2}(2\xi-1)^{2}\nonumber \\
 & \phantom{}\times._{2}F_{1}(\frac{1}{2},D+\frac{3}{2};\frac{3}{2},(\frac{3s}{1+3s})^{2}(2\xi-1)^{2}).
\end{align}

Because it keeps the two-channel competition but discards only the
gap correlations, the ISA approximation is generally closer to the
exact result than ESA, especially for the trace tails and for the
shape distribution at small $D$. However, as $D$ increases, the
exact coupling between scale and shape becomes more important, and
the ISA approximation begins to deviate visibly in the bulk. 

The distinction between these two approximations is physically instructive.
The ESA removes essentially all information about the fluctuation
of the gap ratio and is therefore best viewed as a minimal analytic
model. The ISA goes one step further by allowing the two virtual transition
channels to fluctuate independently, but it still misses the fact
that the two channels share the same middle level and therefore inherit
correlated spacing statistics from the underlying GUE spectrum. In
this sense, the difference between the ISA and the exact solution
measures the importance of adjacent-gap correlations for QGT eigenvalue
statistics.

The comparison in Fig.~\ref{fig:n3} illustrates the roles of the two-channel
geometry and the adjacent-gap correlations. Panels $(a,c)$ show the
rescaled trace distribution $Dp_{S}(s/D)$, while panels $(b,d)$
show the shape distribution $p_{\xi}(\xi)$. We display both a small
parameter dimension, $D=2$, and a large one, $D=50$. The Monte Carlo
data agree with the exact finite-$N$ formula, confirming the validity
of the joint scale-shape distribution derived above. The independent-spacing
approximation (ISA) captures the qualitative behavior of both the
trace and the shape distributions, because it retains the competition
between the two virtual transition channels. Its remaining discrepancy
from the exact result, especially in the shape sector, measures the
effect of the correlated adjacent spacings of the $3\times3$ GUE
spectrum.

The exact $N=3$ solution therefore clarifies which ingredients are
essential for QGT eigenvalue statistics. The two-channel structure
is responsible for the emergence of shape fluctuations, while the
correlated adjacent gaps determine the quantitative correction to
simpler gap approximations. These observations suggest that a general
finite-$N$ theory should retain the scale-shape structure of the
spectrum while incorporating the universal constraints visible in
the solvable cases. We now use this perspective to construct an approximate
model for arbitrary finite $N$ and $D$.

\section{model for $p_{S,\xi}(s,\xi)$ for arbitrary $N$ and $D$}
\begin{figure}
\includegraphics[width=\columnwidth]{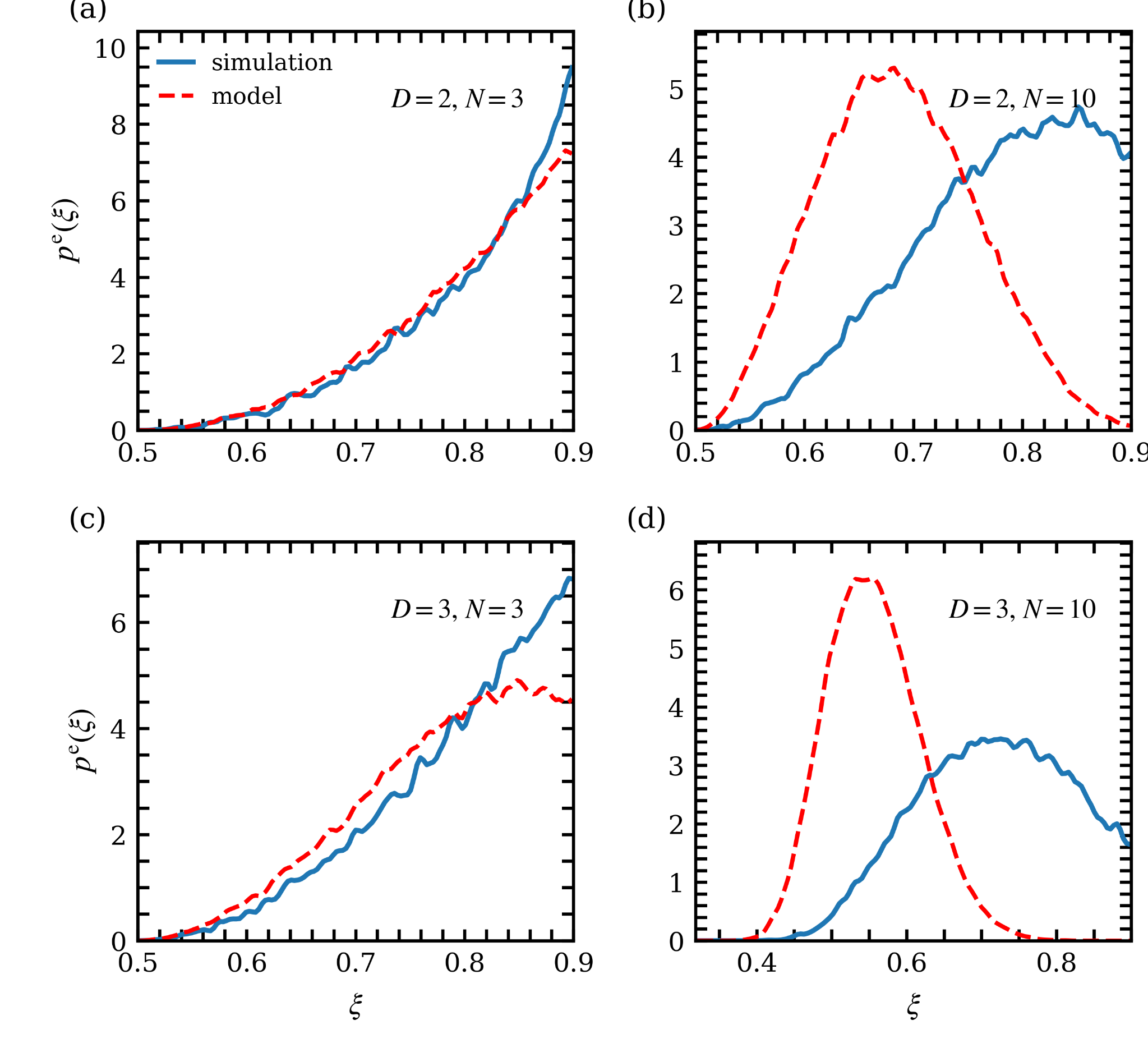}
\begin{turnpage}
\caption{\label{fig:finiteN} Finite-$N$ shape statistics of the QGT. The
distribution of $\xi=\lambda_{\max}/\mathrm{tr}\,G$ is shown for
$(D,N)=(2,3),(2,10),(3,3),(3,10)$. Blue curves are full random-Hamiltonian
simulations, and red dashed curves are the equal-spacing Wishart shape
model.}
\end{turnpage}
\end{figure}

For general $N$ and $D$, obtaining an exact closed-form joint eigenvalue
distribution becomes increasingly difficult. The solvable cases above
nevertheless identify robust structural constraints: the QGT spectrum
admits a scale-shape decomposition, the scale sector is constrained
by the total number of fluctuation channels and by near-degeneracy
tails, and the shape sector is naturally approximated by a finite-rank
Wishart structure when energy-denominator fluctuations are suppressed.
We therefore construct a minimal finite-$N$ scale-shape model that
preserves these features.

\subsection{Approximation of Scale {\normalsize\textmd{$p_{S}(s)$}}}

\subsubsection{Exact asymptotic input for the scale sector}
The scale variable $S=\mathrm{tr}\,G$ receives contributions from
all $D(N-1)$ transition amplitudes. Small $S$ requires all these
amplitudes to be simultaneously suppressed, giving $p_{S}(s)\sim s^{D(N-1)-1}$.
Large $S$ is dominated by rare near-degeneracies in the spectrum,
giving the universal tail $p_{S}(s)\sim s^{-5/2}$. These two exponents
will be used as the main constraints on the scale ansatz.

The mean scale can also be evaluated analytically. In Appendix D,
we show that
\begin{equation}
\mathbb{E}[S]=\frac{(N-1)(D+(D-1)r^{2})}{2(r^{2}+1)^{2}}.
\end{equation}

\subsubsection{Beta-prime ansatz for $p_{S}(s)$}
Among simple positive distributions, the beta-prime family provides
the minimal interpolation between independent power laws at the head
and at the tail. It is therefore the natural first candidate for $p_{S}(s)$.
We write the beta-prime density as
\begin{equation}
p_{beta-prime}(s)=\frac{(\frac{s}{s^{*}})^{\alpha-1}(1+\frac{s}{s^{*}})^{-\alpha-\gamma}}{s^{*}B(\alpha,\gamma)}.
\end{equation}

Matching the two asymptotic exponents and the mean scale, this gives
\begin{equation}
p_{S}(s)\approx\frac{4(4s)^{D(N-1)-1}(1+4s)^{-D(N-1)-\frac{5}{2}}}{B(D(N-1),\frac{3}{2})}
\end{equation}

\subsection{Leading-order shape ansatz from the ESA approximation}
The shape sector is modeled by suppressing fluctuations of the energy
denominators. In this limit, the QGT becomes a Gram matrix formed
from random transition vectors. Its eigenvalue statistics are therefore
described by a finite-rank Wishart ensemble \citep{potters2020first}.
Physically, this approximation keeps the random orientation of the
transition channels in parameter space while treating the energy denominators
as an overall scale:
\begin{equation}
G^{(n)}\equiv\frac{1}{\Delta^{2}}X^{(n)}X^{(n)\dagger},X_{\alpha m}\equiv\langle n|H_{\alpha}|m\rangle.
\end{equation}

Here $X$ is a $D\times(N-1)$ matrix, so $X^{(n)}X^{(n)\dagger}$
forms a Wishart matrix. The joint distribution of non-zero eigenvalues is given by 
\begin{align}
p_{\lambda}(\lambda_{1},...,\lambda_{r}|\Delta) & \approx C_{r,\rho}(N\Delta^{2})^{D(N-1)}\mathrm{Exp}(-N\Delta^{2}\sum_{i=1}^{r}\lambda_{i})\nonumber \\
 & \times\prod_{i=1}^{r}\lambda_{i}^{\rho}\prod_{1\leq i<j\leq r}(\lambda_{i}-\lambda_{j})^{2}
\end{align}
where $r=\mathrm{min}(D,N-1)$, $\rho=|D-(N-1)|$, $C_{r,\rho}=1/\prod_{j=1}^{r}\Gamma(j+1)\Gamma(\rho+j)$.
This leads to a concise representation of $p_{\xi}(\xi|s)$:
\begin{equation}
p_{\xi}(\xi|s)\approx\Gamma(D(N-1))C_{r,\rho}\delta_{1,\sum\xi_{i}}\prod_{i=1}^{r}\xi_{i}^{\rho}\prod_{1\leq i<j\leq r}(\xi_{i}-\xi_{j})^{2}.
\end{equation}
The right-hand side is independent of $S$, so the shape sector decouples
from the scale sector at leading ESA order.

The quality of this leading shape ansatz is tested in Fig.~\ref{fig:finiteN}.
We compare the normalized largest eigenvalue $\xi=\lambda_{\max}/\mathrm{tr}\,G$,
obtained from the full random-Hamiltonian simulation with the equal-spacing
Wishart shape model. For $N=3$, the shape distribution is only weakly
modified when the number of parameters is changed from $D=2$ to $D=3$.
This indicates that the normalized QGT spectrum is mainly controlled
by the finite-rank structure of the Gram matrix, rather than by the
precise value of the parameter dimension. The Wishart shape model
therefore provides a useful baseline for arbitrary finite $N$ and
$D$.

The discrepancy becomes more visible for larger $N$. This is expected,
because the ESA suppresses the nonuniformity and correlations of the
energy denominators. As $N$ grows, the QGT receives contributions
from many levels with different spacings, and the shape sector becomes
increasingly sensitive to these denominator fluctuations. Thus Figure
(3) should not be interpreted as an exact finite-$N$ result, but
as a test of the effective finite-rank Wishart structure that remains
after the scale fluctuations have been separated.

\section{resonant-disordered realization}
To connect the finite-dimensional random-Hamiltonian model with a
microscopic disordered system, we consider a low-dimensional effective
Hamiltonian emerging from a resonant subspace. Specifically, we study
a two-dimensional Anderson lattice with on-site disorder and finite-range
complex hopping, extended by external parameters $y=(y_{1},\ldots,y_{D})$
that control additional random hopping layers. The full Hamiltonian is
\begin{equation}
H(\mathbf{y})=\sum_{i}\epsilon_{i}|i\rangle\langle i|+\sum_{ij}(J_{0})_{ij}+y^{\mu}(J_{\mu})_{ij}|i\rangle\langle j|,
\end{equation}
where the on-site energies $\{\epsilon_{i}\}$ are independent and
identically distributed random variables with disorder strength $W$,
and the hopping amplitudes $\{(J_{0})_{ij},(J_{\mu})_{ij}\}$ are
independent random hopping layers. The parameters $y_{\mu}$ probe
the local quantum geometry of the effective resonant subspace.
\begin{figure}
\subfloat{\includegraphics[width=\columnwidth]{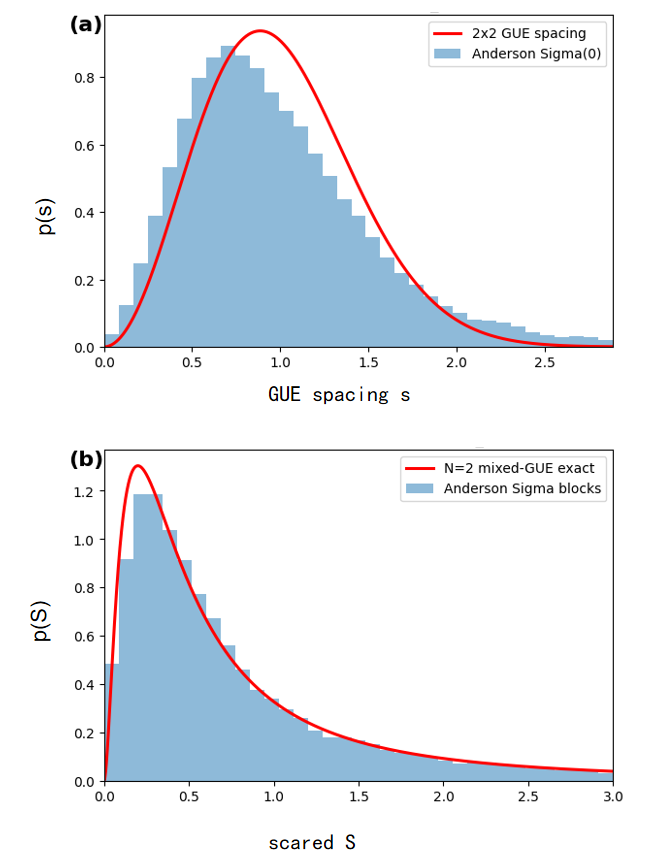}}
\caption{\label{fig:resonant}Numerical resonant-disordered realization. $(a)$ Mean-normalized
level spacing $s=\Delta/\mathbb{E}[\Delta]$ of the two-site self-energy
$\Sigma(0)=H_{{\rm eff}}(0)-PHP$. $(b)$ Window-normalized QGT trace
distribution for $S\le3$. Red curves denote the $2\times2$ GUE spacing
distribution and the exact $N=2$ mixed-GUE trace distribution.}
\end{figure}

In the strong-disorder regime $W\gg|J_{ij}|$, most eigenstates are
localized \citep{anderson1958absence,frohlich1985constructive}. Nevertheless,
rare near-degenerate sites or clusters satisfying $|\epsilon_{i}-\epsilon_{j}|\lesssim|J_{ij}|$
can hybridize and form resonant eigenstates \citep{johri2012singular,bhatt2012rare,ammari2024anderson}.
The relevant low-energy dynamics near such a resonance is therefore
described not by the full Hilbert space, but by a finite resonant
subspace. This provides a microscopic route to the finite-$N$ random
Hamiltonians studied above.

Let $P$ be the projector onto a chosen resonant subspace and $Q=I-P$.
Eliminating the non-resonant sector gives the energy-dependent effective
Hamiltonian
\begin{equation}
H_{{\rm eff}}(E)=PHP+PHQ(E-QHQ)^{-1}QHP.
\end{equation}

The second term describes virtual transitions through the non-resonant
localized states. If we decompose $QHQ=D+V$, where $D$ contains
the localized onsite energies and $V$ contains the residual couplings
among non-resonant states, then
\begin{equation}
(E-QHQ)^{-1}=G\sum_{n=0}^{\infty}(VG)^{n},\qquad G=(E-D)^{-1}.
\end{equation}
Therefore
\begin{equation}
H_{{\rm eff}}(E)=PHP+\sum_{n=0}^{\infty}(PHQ)G(VG)^{n}(QHP).
\end{equation}

This expression shows that the matrix elements inside the resonant
subspace receive contributions from many virtual hopping paths. When
sufficiently many paths contribute, their random amplitudes, phases,
and energy denominators tend to wash out microscopic details. After
an irrelevant scalar energy shift is subtracted, the effective resonant
block can therefore be modeled by a Wigner-type Hermitian random matrix.
In the absence of additional antiunitary or chiral constraints, the
corresponding minimal ensemble is the GUE:
\begin{equation}
\Sigma\equiv H_{{\rm eff}}(E)-PHP
\end{equation}

The parameter dependence of the microscopic hopping amplitudes induces
a corresponding parameter dependence of the effective resonant Hamiltonian.
Expanding around a reference point $\mathbf{y}_{0}$, we obtain
\begin{equation}
\Sigma=\Sigma_{0}+\sum_{\mu=1}^{D}y_{\mu}\left.\partial_{\mu}\Sigma\right|_{\mathbf{y}=0}+O(|\mathbf{y}-\mathbf{y}_{0}|^{2}).
\end{equation}

Keeping only the leading local dependence gives
\begin{equation}
\Sigma(y)=\Sigma_{0}+\sum_{\mu=1}^{D}y_{\mu}\,\partial_{\mu}\Sigma\big|_{y=0}.
\end{equation}
which is precisely the linear mixed-GUE structure studied in the preceding
sections. In this sense, the random QGT eigenvalue statistics derived
above can be viewed as a universal local description of quantum geometry
near resonant disordered eigenstates.

In the numerical realization shown in Fig.~\ref{fig:resonant}, we use $D=3$ external
parameters and choose the hopping layers as
\begin{equation}
J_{0ij}\stackrel{d}{=}J_{\mu ij}=\begin{cases}
t\mathrm{exp}(-\frac{|i-j|}{l})(\frac{a_{\mu ij}+ib_{\mu ij}}{2}) & |i-j|\leq R_{c}\\
0 & |i-j|>R_{c}
\end{cases},
\end{equation}
with Hermiticity imposed by$J_{ji}^{(\mu)}=J_{ij}^{(\mu)*}$. The
onsite energies are uniformly distributed in $[-W/2,W/2]$. 

For each selected pair, we compute the self-energy $\Sigma(\mathbf{y})=H_{PQ}(\mathbf{y})\left[E-H_{QQ}(\mathbf{y})\right]^{-1}H_{QP}(\mathbf{y}),$
and remove its scalar part before extracting the level spacing. Figure
$(4a)$ shows the spacing distribution of $\Sigma(0)$, with $s=\Delta/\mathbb{E}[\Delta],\qquad\Delta=E_{2}-E_{1}.$
The distribution exhibits clear level repulsion and is close to the
$2\times2$ GUE Wigner surmise. The remaining small deviation is attributed
to finite-connectivity covariance anisotropy: in a finite-range lattice,
the diagonal self-energy difference and the off-diagonal self-energy
do not have exactly equal variances. Fig.~\ref{fig:resonant}(b) shows the corresponding
QGT trace, $S=\mathrm{tr}\,G=\sum_{\mu=1}^{D}\frac{|\langle1|\partial_{\mu}\Sigma|2\rangle|^{2}}{(E_{2}-E_{1})^{2}}.$
The agreement with the exact $N=2$ result shows that the virtual-path
self-energy of a two-site resonant block reproduces the leading mixed-GUE
quantum-geometric statistics.

This realization also clarifies why the cases $N=2$ and $N=3$ are
especially relevant. In a strongly disordered system, large resonant
clusters are rare, while two-level and three-level resonances are
the most common nontrivial hybridized structures. The exact finite\nobreakdash-N
results obtained in this work therefore provide solvable benchmarks
for random quantum geometry and effective descriptions of the QGT
associated with rare resonant states in disordered systems. In Appendix
E, we show that the probability of finding an $N$-site resonant block
decreases rapidly with $N$.

\section{conclusion and discussion}
In this work, we developed an arbitrary-$D$ theory for the eigenvalue
statistics of the QGT in parameter-dependent random Hamiltonians.
The central point is that the QGT should be treated not only through
scalar observables such as the trace, but through its full nonzero
spectrum, which resolves the scale and shape of the QGT. By exploiting
the $U(D)$ invariance of the QGT distribution, we reduced the problem
to the statistics of its eigenvalues, which provide a natural description
of the scale, anisotropy, and effective rank of the QGT. We obtained
exact results for the first two nontrivial cases, $N=2$ and $N=3$.
In particular, the $N=3$ case gives the first exact finite-$N$ description
of nontrivial shape fluctuations.

Motivated by these exact solutions, we further proposed a unified
model for arbitrary finite $N$ and $D$, based on a scale-shape decomposition
of the eigenvalue distribution. The model correctly captures the universal
head and tail behavior of the trace and gives a simple approximation
for the shape sector. Numerical simulations show good agreement with
the exact formulas and support the general model. The resonant-disordered
construction discussed above gives a simple physical interpretation
of the finite-dimensional mixed-GUE ensemble. Our results provide
a useful non-asymptotic framework for random quantum geometry and
offer a basis for future studies of disordered, chaotic, and finite-dimensional
quantum systems. Possible extensions include other random-matrix symmetry
classes, non-Hermitian Hamiltonians, and many-body settings where
QGT eigenvalues may diagnose the structure of quantum-state sensitivity
in high-dimensional parameter spaces.

\medskip{}

\begin{acknowledgments}
This work was financially supported by the National Key R\&D Program
of the MOST of China (Grant No. 2024YFA1611300), the National Natural
Science Foundation of China (Grant No. 12574059), HFIPS Director's
Fund (Grant No. BJPY2023B05), Anhui Provincial Major S\&T Project
(s202305a12020005) and the Basic Research Program of the Chinese Academy
of Sciences Based on Major Scientific Infrastructures (Grant No. JZHKYPT-2021-08)
and the High Magnetic Field Laboratory of Anhui Province under contract
No. AHHM-FX-2020-02.
\end{acknowledgments}

\bibliography{QGTEigenvalues}

\appendix

\section{proof of the $U(D)$ invariance}

In this appendix we give a more explicit proof of the statistical
$U(D)$ invariance used in the main text. For any $U\in U(D)$ and
any Borel set $A\subset\mathrm{Herm}(D)$, the statement is
\begin{equation}
\mathrm{Pr}\!\left(G^{(n)}\in A\right)=\mathrm{Pr}\!\left(UG^{(n)}U^{\dagger}\in A\right),
\end{equation}
where $\mathrm{Herm}(D)$ means the set of $D\times D$ Hermitian
matrices.

Let $V$ be a unitary matrix that diagonalizes $H_{0}$,
\begin{equation}
V^{\dagger}H_{0}V=\mathrm{diag}(E_{1},\ldots,E_{N}).
\end{equation}

Since $H_{\alpha}$ is independent of $H_{0}$, and since the GUE
ensemble is invariant under unitary conjugation, the rotated matrices
\begin{equation}
\widetilde{H}_{\alpha}=V^{\dagger}H_{\alpha}V,\qquad\alpha=1,\ldots,D,
\end{equation}
remain mutually independent GUE matrices after conditioning on $H_{0}$.
Their conditional distribution does not depend on the eigenvectors
of $H_{0}$.

For each $m\neq n$, define the transition vector
\begin{equation}
(X_{m})_{\alpha}=\frac{(\widetilde{H}_{\alpha})_{nm}}{E_{m}-E_{n}}.
\end{equation}

Then the QGT can be written as the Gram matrix
\begin{equation}
G^{(n)}=\sum_{m\neq n}X_{m}X_{m}^{\dagger}.
\end{equation}

Conditioned on $H_{0}$, the vectors $X_{m}$ are independent centered
complex Gaussian vectors with covariance
\begin{equation}
\mathbb{E}\!\left[(X_{m})_{\alpha}(X_{m'})_{\beta}^{*}\,|\,H_{0}\right]=\frac{\delta_{mm'}\delta_{\alpha\beta}}{N(E_{m}-E_{n})^{2}}.
\end{equation}

The covariance is proportional to the identity in the parameter indices.
Therefore, for every $U\in U(D)$,
\begin{equation}
(UX_{m})_{m\neq n}\stackrel{d}{=}(X_{m})_{m\neq n}
\end{equation}

after conditioning on $H_{0}$, where $\overset{d}{=}$ denotes equality
in distribution. It follows that
\begin{equation}
UG^{(n)}U^{\dagger}=\sum_{m\neq n}(UX_{m})(UX_{m})^{\dagger}\stackrel{d}{=}G^{(n)}
\end{equation}

under the same conditioning. Averaging over $H_{0}$ preserves the
equality in distribution and gives the desired $\ensuremath{U(D)}$
invariance.

Equivalently, the QGT ensemble is invariant under unitary conjugation
in parameter space. This implies that the nontrivial invariant information
is contained in the eigenvalue statistics of $G^{(n)}$.

\section{distribution of the scale variable for $\ensuremath{N=2}$}

We now derive the distribution of the only nonzero QGT eigenvalue
for $N=2$. In this case the scale variable is simply
\begin{equation}
S=\lambda_{\max}=\frac{K}{U},
\end{equation}
where
\begin{equation}
K=\sum_{\alpha=1}^{D}|\langle1|H_{\alpha}|2\rangle|^{2},\qquad U=(E_{2}-E_{1})^{2}.
\end{equation}

The two random variables $K$ and $U$ are independent because the
transition matrix elements are independent of the eigenvalues of $H_{0}$.

\subsection{The Density of $K,U$}

We first determine the distribution of $K$. For each parameter direction,
the off-diagonal matrix element $\langle1|H_{\alpha}|2\rangle$ is
a complex Gaussian variable. Hence $|\langle1|H_{\alpha}|2\rangle|^{2}$
is exponentially distributed. Its characteristic function is
\begin{equation}
\chi(t)=\mathbb{E}\!\left[e^{it|\langle1|H_{\alpha}|2\rangle|^{2}}\right]=\frac{2}{2-it}.
\end{equation}

Since $\ensuremath{K}$ is the sum of $D$ independent copies of this
variable, its characteristic function is
\begin{equation}
\chi_{K}(t)=\left(1-\frac{it}{2}\right)^{-D}.
\end{equation}

Therefore
\begin{equation}
p_{K}(x)=\frac{2^{D}x^{D-1}e^{-2x}}{(D-1)!}.
\end{equation}

Next, let $\Delta=E_{2}-E_{1}$. For a $2\times2$ GUE matrix, the
two level spacing distribution is \citep{berry2020quantum}
\begin{equation}
p_{\Delta}(y)=\sqrt{\frac{2}{\pi}}\,y^{2}e^{-y^{2}/2}.
\end{equation}

Thus $U=\Delta^{2}$ has density
\begin{equation}
p_{U}(u)=\frac{1}{2^{3/2}\Gamma(3/2)}u^{1/2}e^{-u/2}.
\end{equation}

\subsection{The Density of $S$}

Since $S=K/U$, we use the change of variables$K=su$. The Jacobian
gives
\begin{equation}
p_{S}(s)=\int_{0}^{\infty}u\,p_{K}(su)p_{U}(u)\,du.
\end{equation}

Evaluating the integral yields
\begin{equation}
p_{S}(s)=\frac{4^{D}}{B(D,3/2)}s^{D-1}(1+4s)^{-(D+3/2)}.
\end{equation}

Thus the scale variable follows a beta-prime distribution. This completes
the derivation of the $N=2$ result used in the main text.

\section{eigenvalue distribution for $\ensuremath{N=3}$}

This appendix gives the derivation of the exact $\ensuremath{N=3}$
eigenvalue distribution. We focus on the middle level. 

\subsection{The Expression of Eigenvalues}

The QGT receives two contributions, one from each neighboring level,
and can be written as
\begin{equation}
G=\mathcal{A}\mathcal{A}^{\dagger}+\mathcal{B}\mathcal{B}^{\dagger},
\end{equation}
where
\begin{equation}
\mathcal{A}=\frac{\omega^{(1)}}{\Delta_{1}},\qquad\mathcal{A}=\frac{\omega^{(3)}}{\Delta_{3}}.
\end{equation}

Let $\Phi=(A,B)$ be the $D\times2$ matrix formed by the two transition
vectors. Then
\begin{equation}
G=\Phi\Phi^{\dagger}.
\end{equation}

The nonzero eigenvalues of $G$ coincide with those of the $2\times2$
matrix $\Phi^{\dagger}\Phi$. Therefore
\begin{equation}
\Phi^{\dagger}\Phi=\begin{pmatrix}A^{\dagger}A & A^{\dagger}B\\
B^{\dagger}A & B^{\dagger}B
\end{pmatrix}.
\end{equation}

Solving this two-dimensional eigenvalue problem gives
\begin{equation}
\lambda_{\pm}=\frac{1}{2}\left[X^{(1)}+X^{(3)}\pm\sqrt{(X^{(1)}-X^{(3)})^{2}+4X^{(1)}X^{(3)}\phi}\right],
\end{equation}
where
\[
X^{(1)}\equiv\frac{|\omega^{(1)}|^{2}}{\Delta_{1}^{2}},\,X^{(3)}\equiv\frac{|\omega^{(3)}|^{2}}{\Delta_{3}^{2}},
\]
\begin{equation}
u^{(1)}\equiv\frac{\omega^{(1)}}{|\omega^{(1)}|},\,u^{(3)}\equiv\frac{\omega^{(3)}}{|\omega^{(3)}|},\,\phi\equiv|u^{(1)\dagger}u^{(3)}|^{2}.
\end{equation}

Here $X^{(1)}$ and $X^{(3)}$ are the two channel weights, while
$\phi$ measures the relative orientation of the two transition vectors.

\subsection{Conditional density $p_{S,\xi,z}(s,\xi,z|\Delta_{1},\Delta_{3})$}

\subsubsection{\textmd{\textup{Density of $S$ and $z$ at fixed spacings}}}

We first keep the two adjacent spacings $\ensuremath{\Delta_{1}}$and
$\Delta_{3}$ fixed. Under this conditioning, the two channel weights
$X^{(1)}$ and $X^{(3)}$ are independent Gamma variables. It is also
convenient to introduce the relative weight $z=\frac{X^{(1)}}{X^{(1)}+X^{(3)}}.$

The scale variable is $S=X^{(1)}+X^{(3)}$, and the shape variable
is chosen as $\xi=\lambda_{+}/S$. By the same argument used in $N=2$
case,
\begin{align}
p_{X^{(1)}}(x|\Delta_{1}) & =\Delta_{1}^{2D}\frac{3^{D}x^{D-1}\mathrm{Exp}(-3\Delta_{1}^{2}x)}{\Gamma(D)}\\
p_{X^{(3)}}(y|\Delta_{3}) & =\Delta_{3}^{2D}\frac{3^{D}y^{D-1}\mathrm{Exp}(-3\Delta_{3}^{2}y)}{\Gamma(D)}.
\end{align}

Therefore
\begin{align}
p_{S,z}(s,z|\Delta_{1},\Delta_{3}) & =sp_{X^{(1)}}(sz|\Delta_{1})p_{X^{(3)}}(s(1-z)|\Delta_{3})\nonumber \\
 & =(\Delta_{1}\Delta_{3})^{2D}\frac{3^{2D}}{\Gamma^{2}(D)}(z(1-z))^{D-1}s^{2D-1}\nonumber \\
 & \times\mathrm{Exp}(-3(\Delta_{1}^{2}sz+\Delta_{3}^{2}s(1-z))).
\end{align}

Here we have used the conditional independence of $X^{(1)}$ and $X^{(2)}$
at fixed $\Delta_{1}$ and $\Delta_{3}$.

\subsubsection{Density of the angular variable $\phi$ }

The angular variable $\phi$ has a simple distribution. By unitary
invariance, one may fix one of the two unit vectors and regard $\phi$
as the squared modulus of one component of a random unit vector in
$\mathbb{C}^{D}$. Equivalently, after fixing $u^{(3)}$, we may write
\begin{equation}
\phi=|u^{(1)\dagger}u^{(3)}|^{2}=|u_{1}^{(1)}|^{2}=\frac{|\omega_{1}^{(1)}|^{2}}{\sum_{i=1}^{D}|\omega_{i}^{(1)}|^{2}}.
\end{equation}

Let $X=|\omega_{1}^{(1)}|^{2}$, $Y=\sum_{i=2}^{D}|\omega_{i}^{(1)}|^{2}$,then
\begin{align}
X & \sim\mathrm{Exp}(3)\sim\mathrm{Gamma}(1,3)\nonumber \\
Y & \sim\mathrm{Gamma}(D-1,3).
\end{align}

And two variables are independent. Hence
\begin{equation}
\phi=\frac{X}{X+Y}\sim\mathrm{Beta(1,D-1)}.
\end{equation}

Indeed, setting $t=X+Y$ gives $X=t\phi,Y=t(1-\phi)$, so
\begin{align}
p_{t,\phi}(t,\phi) & =tp_{X}(\phi t)p_{Y}(t(1-\phi))\nonumber \\
 & =\frac{\lambda^{D}}{(D-2)!}(1-\phi)^{D-2}t^{D-1}e^{-\lambda t},
\end{align}
with $\lambda=3$. Integrate over $t$ gives

\begin{align}
p_{\phi}(\phi) & =\int p_{t,\phi}(t,\phi)dt\nonumber \\
 & =(D-1)(1-\phi)^{D-2}.
\end{align}

\subsubsection{\textmd{\textup{Density of $\xi$ at fixed $z$}}}

We now transform from $\phi$ to $\xi$at fixed $z$. From the eigenvalue
formula,
\begin{equation}
\phi=\frac{z(1-z)-(\xi-\xi^{2})}{z(1-z)},\qquad 1-\phi=\frac{(\xi-\xi^{2})}{z(1-z)},
\end{equation}

and $\frac{d\phi}{d\xi}=\frac{2\xi-1}{z(1-z)}.$ It follows that
\begin{align}
p_{\xi,z}(\xi,z) & =p_{\phi}(\phi)|\frac{d\phi}{d\xi}|\nonumber \\
 & =(D-1)[\xi(1-\xi)]^{D-2}(2\xi-1)[\frac{1}{z(1-z)}]^{D-1}.
\end{align}

\subsubsection{\textmd{\textup{Conditional joint density of $S$ and $\xi$}}}

Combining the previous results gives
\begin{align}
 & \phantom{}p_{S,\xi}(s,\xi|\Delta_{1},\Delta_{3})\nonumber \\
 & =\int_{1-\xi}^{\xi}p_{s,z}(s,z|\Delta_{1},\Delta_{3})p_{\xi}(\xi,z)dz\nonumber \\
 & =\frac{(D-1)(\Delta_{1}\Delta_{3})^{2D}3^{2D}s^{2D-1}}{\Gamma(D)^{2}}[\xi(1-\xi)]^{D-2}\nonumber \\
 & \times(2\xi-1)\int_{1-\xi}^{\xi}\mathrm{Exp}(-3(\Delta_{1}^{2}sz+\Delta_{3}^{2}s(1-z)))dz.
\end{align}

Notice that $\xi=\frac{1}{2}(1+\sqrt{1-4z(1-z)(1-\phi)})$, so $\xi\in[max(z,1-z),1]$.
Integrate respect to $z$ and get 
\begin{align}
 & \phantom{}\int_{1-\xi}^{\xi}e^{-3(\Delta_{1}^{2}sz+\Delta_{3}^{2}s(1-z))}dz\nonumber \\
 & =e^{-3\Delta_{3}^{2}s}\int_{1-\xi}^{\xi}e^{-3(\Delta_{1}^{2}-\Delta_{3}^{2})sz)}dz\nonumber \\
 & =\frac{e^{-3\Delta_{3}^{2}s}}{3(\Delta_{1}^{2}-\Delta_{3}^{2})s}[\mathrm{Exp}(-3(\Delta_{1}^{2}-\Delta_{3}^{2})s(1-\xi))\nonumber \\
 & -\mathrm{Exp}(-3(\Delta_{1}^{2}-\Delta_{3}^{2})s\xi))].
\end{align}

Hence 
\begin{align}
 & \phantom{}p_{S,\xi}(s,\xi|\Delta_{1},\Delta_{3})\nonumber \\
 & =\frac{(D-1)(\Delta_{1}\Delta_{3})^{2D}3^{2D-1}s^{2D-2}}{\Gamma(D)^{2}(\Delta_{1}^{2}-\Delta_{3}^{2})}\nonumber \\
 & \times[\xi(1-\xi)]^{D-2}(2\xi-1)[\mathrm{Exp}(-3(\Delta_{1}^{2}s+(\Delta_{3}^{2}-\Delta_{1}^{2})s\xi))\nonumber \\
 & -\mathrm{Exp}(-3(\Delta_{1}^{2}s\xi+\Delta_{3}^{2}s(1-\xi))))].
\end{align}

This expression already shows the key difference from the $N=2$ case:
the scale and shape variables are coupled through the two adjacent
level spacings.

\subsection{Averaging over GUE spacings}

The final step is to average over the GUE spacing distribution.

\subsubsection{The density of adjacent spacing}

The eigenvalue distribution of $H_{0}$ is \citep{potters2020first,mehta2004random}.
\begin{align}
p_{E}(E_{1},E_{2},E_{3}) & \propto(E_{1}-E_{2})^{2}(E_{1}-E_{3})^{2}(E_{2}-E_{3})^{2}\nonumber \\
 & \mathrm{Exp}(-\frac{3}{2}(E_{1}^{2}+E_{2}^{2}+E_{3}^{2})).
\end{align}

We define
\begin{equation}
u\equiv\frac{E_{1}+E_{2}+E_{3}}{3}.
\end{equation}

Then
\begin{equation}
E_{1}^{2}+E_{2}^{2}+E_{3}^{2}=3u^{2}+\frac{2}{3}(\Delta_{1}^{2}+\Delta_{3}^{2}+\Delta_{1}\Delta_{3}),
\end{equation}
and hence

\begin{align}
p_{E}(\Delta_{1},\Delta_{3},u) & \propto\Delta_{1}^{2}\Delta_{3}^{2}(\Delta_{1}+\Delta_{3})^{2}.\nonumber \\
 & \times\mathrm{Exp}(-\frac{9}{2}u^{2}-(\Delta_{1}^{2}+\Delta_{3}^{2}+\Delta_{1}\Delta_{3})).
\end{align}

One obtains

\begin{align*}
p_{\Delta}(\Delta_{1},\Delta_{3}) & =\frac{81}{4\sqrt{3}\pi}\Delta_{1}^{2}\Delta_{3}^{2}(\Delta_{1}+\Delta_{3})^{2}\\
 & \times\mathrm{Exp}(-(\Delta_{1}^{2}+\Delta_{3}^{2}+\Delta_{1}\Delta_{3})).
\end{align*}

\subsubsection{Exact joint density$p_{S,\xi}(s,\xi)$}

Introduce $x\equiv\Delta_{1}+\Delta_{3},y\equiv\frac{\Delta_{1}-\Delta_{3}}{\Delta_{1}+\Delta_{3}}\in[-1,1]$,
the Jacobian is 
\begin{align}
|\frac{\partial(\Delta_{1},\Delta_{3})}{\partial(x,y)}| & =\frac{x}{2}.\\
\nonumber 
\end{align}

Substituting the conditional density and integrating over the spacing
distribution, we obtain
\begin{align}
p_{S,\xi}(s,\xi) & =\int p_{S,\xi}(s,\xi|\Delta_{1},\Delta_{3})p_{\Delta}(\Delta_{1},\Delta_{3})d\Delta_{1}d\Delta_{3}\nonumber \\
 & =K_{D}s^{2D-2}[\xi(1-\xi)]^{D-2}(2\xi-1)\nonumber \\
 & \times\int_{0}^{\infty}dx\int_{-1}^{1}dyx^{4D+5}(1-y^{2})^{2D+2}\nonumber \\
 & \times\mathrm{Exp}(-\frac{x^{2}}{4}[3(1+s)+(1+3s)y^{2}])\\
 & \times\frac{\mathrm{sinh}(\frac{3}{2}sx^{2}y(2\xi-1))}{y},
\end{align}
with $K_{D}=\frac{(D-1)3^{2D+3}}{2^{4D+6}\sqrt{3}\pi\Gamma(D)^{2}}.$

Using the expansion
\begin{equation}
\frac{\mathrm{sinh}(\frac{3}{2}sx^{2}y(2\xi-1))}{y}=\sum_{l=0}^{\infty}\frac{(\frac{3}{2}s(2\xi-1))^{2l+1}}{(2l+1)!}x^{4l+2}y^{2l},
\end{equation}
we define
\begin{align}
I_{l}(s,\xi) & \equiv\int_{0}^{\infty}dx\int_{-1}^{1}dyx^{4D+4l+7}(1-y^{2})^{2D+2}y^{2l}.\\
 & \times\mathrm{Exp}(-\frac{x^{2}}{4}[3(1+s)+(1+3s)y^{2}]).
\end{align}

Integrating first over $x$ give
\begin{align}
I_{l} & =\Gamma(2D+2l+4)2^{4D+4l+8}\nonumber \\
 & \times\int_{-1}^{1}dy(1-y^{2})^{2D+2}y^{2l}[3(1+s)+(1+3s)y^{2}]^{-(2D+2l+4)}.
\end{align}
 Using Euler's integral formula \citep{abramowitz1948handbook}, this
becomes
\begin{align}
 & \phantom{}I_{l}\nonumber \\
 & =\Gamma(2D+2l+4)2^{4D+4l+8}[3(1+s)]^{-(2D+2l+4)}B(l+\frac{1}{2},2D+3)\nonumber \\
 & \times{}_{2}F_{1}(2D+2l+4,l+\frac{1}{2};l+2D+\frac{7}{2},-\frac{1+3s}{3(1+s)}).
\end{align}

Finally,
\begin{equation}
p_{S,\xi}(s,\xi)=[\xi(1-\xi)]^{D-2}\sum_{l=0}^{\infty}c_{l}(s)(2\xi-1)^{2l+2},
\end{equation}
where
\begin{align}
c_{l}(s) & =\frac{(D-1)3^{2D+2l+4}2^{2l}\Gamma(2D+2l+4)}{\sqrt{3}\pi\Gamma(D)^{2}(2l+1)!}\nonumber \\
 & \phantom{}\times s^{2D+2l-1}[3(1+s)]^{-(2D+2l+4)}B(l+\frac{1}{2},2D+3)\nonumber \\
 & \phantom{}\times{}_{2}F_{1}(2D+2l+4,l+\frac{1}{2};l+2D+\frac{7}{2},-\frac{1+3s}{3(1+s)}).
\end{align}

\subsection{Approximation Distributions}

We next derive the two approximate distributions used for comparison
in the main text. The equal-spacing approximation is obtained by setting
the two adjacent gaps equal. This removes the fluctuation of the gap
ratio and reduces the problem to a symmetric two-channel geometry.

The independent-spacing approximation keeps the two gaps fluctuating
but neglects their correlation. In this case the exact GUE spacing
density is replaced by a product form. Starting from the conditional
density $p_{S,z}(s,z|\Delta_{1},\Delta_{3})$, we first integrate
over $\ensuremath{\Delta_{1}}$ and $\ensuremath{\Delta_{3}}$, using
\[
p_{\Delta}^{ISA}(\Delta_{1},\Delta_{3})=\frac{2}{\pi}\Delta_{1}^{2}\Delta_{3}^{2}\mathrm{Exp}(-\frac{\Delta_{1}^{2}+\Delta_{3}^{2}}{2}).
\]

Thus
\begin{align}
 & p_{S,\xi}^{ISA}(s,\xi)\nonumber \\
 & =\int p_{S,z}(s,z|\Delta_{1},\Delta_{3})p_{\xi}(\xi,z)p_{\Delta}^{ISA}(\Delta_{1},\Delta_{3})d\Delta_{1}d\Delta_{3}dz.
\end{align}

The Gaussian integrals give
\begin{align}
\int_{0}^{\infty}(\Delta_{1}\Delta_{3})^{2D+2}e^{-3(\Delta_{1}^{2}sz+\Delta_{3}^{2}s(1-z))}\nonumber \\
\mathrm{Exp}(-\frac{\Delta_{1}^{2}+\Delta_{3}^{2}}{2})d\Delta_{1}d\Delta_{3}\nonumber \\
=\frac{\Gamma^{2}(D+\frac{3}{2})}{4(3sz+\frac{1}{2})^{D+\frac{3}{2}}(3s(1-z)+\frac{1}{2})^{D+\frac{3}{2}}}.
\end{align}

We are then left with the integral over the relative weight $z$:
\begin{align*}
p_{S,\xi}^{ISA}(s,\xi) & =K'_{D}s^{2D-1}[\xi(1-\xi)]^{D-2}(2\xi-1)\\
 & \times\int_{1-\xi}^{\xi}\frac{dz}{(3sz+\frac{1}{2})^{D+\frac{3}{2}}(3s(1-z)+\frac{1}{2})^{D+\frac{3}{2}}},
\end{align*}
where $K'_{D}=\frac{(D-1)3^{2D}\Gamma^{2}(D+\frac{3}{2})}{2\pi\Gamma^{2}(D)}.$
To evaluate the remaining integral, define
\begin{equation}
p=D+\frac{3}{2},u=2\xi-1,z=\frac{1+y}{2},c=\frac{1+3s}{2},q=\frac{3s}{2}.
\end{equation}

Then
\begin{align*}
J(s,\xi) & =\int_{1-\xi}^{\xi}\frac{1}{(3sz+\frac{1}{2})^{D+\frac{3}{2}}}\frac{1}{(3s(1-z)+\frac{1}{2})^{D+\frac{3}{2}}}dz\\
 & =uc^{-2p}._{2}F_{1}(\frac{1}{2},p;\frac{3}{2},\frac{q^{2}u^{2}}{c^{2}}).
\end{align*}

Thus
\begin{align}
p_{S,\xi}^{ISA}(s,\xi) & =K'_{D}s^{2D-1}[\xi(1-\xi)]^{D-2}(2\xi-1)^{2}c^{-2p}\nonumber \\
 & \times._{2}F_{1}(\frac{1}{2},p;\frac{3}{2},\frac{q^{2}u^{2}}{c^{2}})\nonumber \\
 & =\frac{(D-1)3^{2D}\Gamma^{2}(D+\frac{3}{2})}{2\pi\Gamma^{2}(D)}s^{2D-1}(\frac{1+3s}{2})^{-(2D+3)}\nonumber \\
 & \times[\xi(1-\xi)]^{D-2}(2\xi-1)^{2}\nonumber \\
 & \times{}_{2}F_{1}(\frac{1}{2},D+\frac{3}{2};\frac{3}{2},(\frac{3s}{1+3s})^{2}(2\xi-1)^{2}).
\end{align}

\section{finite-$N$ scale-shape model}

In this appendix we collect the ingredients used to construct the
finite-$N$, arbitrary-$D$ model in the main text. The model is based
on two observations. First, the scale variable $S=\mathrm{tr}\,G$
is strongly constrained by its small-$S$ behavior and large-$S$
tail. Second, when the fluctuations of the energy denominators are
suppressed, the shape sector reduces to the eigenvalue statistics
of a finite-rank Wishart matrix.

\subsection{Tail and Head Behavior}

We first discuss the asymptotic behavior of the scale density. Berry
and Shukla use codimention method to discuss head and tail behavior
of trace when $D=3$ \citep{berry2020quantum}.Here we extend this
argument to arbitrary finite $D$.

For large $S$, the dominant events are near-degeneracies of $H_{0}$.
If a pair of levels $E_{n}$ and $E_{m}$ becomes close, then
\[
S\sim\frac{K^{(n,m)}}{\Delta_{nm}^{2}},
\]
where$K^{(n,m)}\equiv\sum_{\alpha=1}^{D}\langle n|\partial_{\alpha}H|m\rangle\langle m|\partial_{\alpha}H|n\rangle$
is the transition strength and $\Delta_{nm}=|E_{n}-E_{m}|$. 

The small-spacing behavior of the GUE spectrum gives $p(\Delta)\sim\Delta^{2}$
\citep{le2007nearest}, Therefore,
\begin{align}
\mathrm{Pr}(S>s) & =\mathrm{Pr}(\frac{K^{(n,m)}}{\Delta_{nm}^{2}}>s)\nonumber \\
 & =\mathrm{Pr}(\Delta_{nm}<\sqrt{\frac{K^{(n,m)}}{s}})\nonumber \\
 & =\int\mathrm{Pr}(\Delta_{nm}<\sqrt{\frac{k}{s}})p_{K^{(n,m)}}(k)dk.
\end{align}

For fixed $k$, 
\begin{align}
\mathrm{Pr}(\Delta_{nm}<\sqrt{\frac{k}{s}}) & \sim\int_{0}^{\sqrt{\frac{k}{s}}}\Delta^{2}d\Delta\nonumber \\
 & =(\frac{k}{s})^{\frac{3}{2}}.
\end{align}

This leads to
\begin{align}
\mathrm{Pr}(S>s) & \sim\frac{1}{s^{\frac{3}{2}}}\int k^{\frac{3}{2}}p_{K^{(n,m)}}(k)dk\\
 & =\frac{1}{s^{\frac{3}{2}}}\mathbb{E}[(K^{(n,m)})^{\frac{3}{2}}].
\end{align}
Since $K^{(n,m)}$ follows a Gamma distribution, the moment $\mathbb{E}[(K^{(n,m)})^{\frac{3}{2}}]$
is finite. Differentiating with respect to $s$ gives the large-$S$
tail
\[
p_{S}(s)\sim s^{-5/2}.
\]

The same argument applies to individual matrix elements of the QGT,
with only the prefactor modified.

For head behavior, recall that
\begin{equation}
p_{K^{(n,m)}}(k)\sim k^{D-1}.
\end{equation}

Conditioned on each $\Delta_{nm}^{-2}$,
\begin{equation}
p_{\Delta_{nm}^{-2}K^{(n,m)}}(k)\sim\Delta_{nm}^{2D}k^{D-1}.
\end{equation}

Convolution method gives
\begin{equation}
p_{S}(s|\Delta)\sim\prod_{m=1,m\not=n}^{N}\Delta_{nm}^{2D}s^{D(N-1)-1}.
\end{equation}

Averaging over the GUE spacings gives
\begin{equation}
p_{S}(s)\sim s^{D(N-1)-1}\mathbb{E}(\prod_{m=1,m\not=n}^{N}\Delta_{nm}^{D}).
\end{equation}

$\mathbb{E}(\prod_{m=1,m\not=n}^{N}\Delta_{nm}^{D})$ is clearly finite
by the regularity of the GUE joint eigenvalue density. Hence $p_{S}(s)\sim s^{D(N-1)-1}$.

For $D=3$, these result is consistent with the result of Berry and
Shukla.

\subsection{Mean value of the QGT trace}

We also need the mean value of the trace. It is useful to introduce
hyperspherical coordinates in parameter space. This calculation applies
at a general point in parameter space.

Recall the Hamiltonian is
\begin{equation}
H=H_{0}+\sum_{i=1}^{D}y_{i}H_{i}.
\end{equation}

Introduce hyperspherical coordinates by
\begin{align}
y_{1} & =r\mathrm{cos}(\phi_{2})\nonumber \\
y_{2} & =r\mathrm{sin}(\phi_{2})\mathrm{cos}(\phi_{3})\nonumber \\
 & \vdots\nonumber \\
y_{k} & =r\prod_{i=2}^{k}\mathrm{sin}(\phi_{i})\mathrm{cos}(\phi_{k+1})\nonumber \\
 & \vdots\nonumber \\
y_{D-1} & =r\prod_{i=2}^{D-1}\mathrm{sin}(\phi_{i})\mathrm{cos}(\phi_{D})\nonumber \\
y_{D} & =r\prod_{i=2}^{D}\mathrm{sin}(\phi_{i}).
\end{align}

Let
\begin{equation}
u(r,\phi)=\frac{1}{r}(y_{1},...,y_{D})^{T},\overrightarrow{H}=(H_{1},...,H_{D})^{T}.
\end{equation}

We rewrite $H$ as $H=H_{0}+ru^{T}\cdot\overrightarrow{H}.$

Define
\begin{align}
P_{k} & \equiv\begin{cases}
\prod_{k=2}^{k}\mathrm{sin}(\phi_{k}) & k>1\\
1 & k=0,1
\end{cases}\\
\mathrm{cos}(\phi_{D+1}) & \equiv1,\phi_{1}\equiv r.
\end{align}

Then
\begin{equation}
y_{k}=r\prod_{i=2}^{k}\mathrm{sin}(\phi_{i})\mathrm{cos}(\phi_{k+1})=rP_{k}\mathrm{cos}(\phi_{k+1}).
\end{equation}

Now we have to consider $\partial_{\phi_{k}}H,\,k\in[D]$. We will
 show that $\{\partial_{\phi_{k}}H,\,k\in[D]\}$ are centered independent
GUE matrices. Their variance are $\{\frac{1}{N},\,\frac{r^{2}P_{i-1}^{2}}{N}$,
$k\in[D]\}$. 

This fact can be proved by several steps.

First,we expand 
\begin{align}
\partial_{\phi_{1}}H & =u^{T}\cdot\overrightarrow{H}\\
\partial_{\phi_{k}}H & =r\frac{\partial u^{T}}{\partial\phi_{k}}\cdot\overrightarrow{H}.
\end{align}

Define
\begin{equation}
\frac{\partial u^{T}}{\partial\phi_{1}}\equiv u^{T}.
\end{equation}

The covariance is 
\begin{align}
\mathbb{E}[\partial_{\phi_{k}}H\partial_{\phi_{s}}H] & =r^{2-\delta_{k1}-\delta_{si}}\sum_{i=1}^{D}\frac{\partial u_{i}^{T}}{\partial\phi_{k}}\frac{\partial u_{i}^{T}}{\partial\phi_{s}}\mathbb{E}[H_{i}H_{j}]\nonumber \\
 & =\frac{r^{2-\delta_{k1}-\delta_{si}}}{N}\frac{\partial u^{T}}{\partial\phi_{k}}\cdot\frac{\partial u^{T}}{\partial\phi_{s}}.
\end{align}

A direct differentiation gives
\begin{equation}
\frac{\partial u_{i}^{T}}{\partial\phi_{k}}=\begin{cases}
0 & i<k-1\\
-P_{k} & i=k-1\\
P_{i}cot(\phi_{k})cos(\phi_{i+1}) & i>k-1
\end{cases}.
\end{equation}

From this expression one obtains the orthogonality relation

\begin{equation}
\frac{\partial u^{T}}{\partial\phi_{k}}\cdot\frac{\partial u^{T}}{\partial\phi_{s}}=P_{k-1}^{2}\delta_{ks}
\end{equation}

Indeed, for $k<s$, 

\begin{align}
\frac{\partial u^{T}}{\partial\phi_{k}}\cdot\frac{\partial u^{T}}{\partial\phi_{s}} & =\mathrm{cot}(\phi_{k})[-P_{s-1}^{2}\mathrm{cot}(\phi_{k})\mathrm{cos}(\phi_{s})\mathrm{sin}(\phi_{s})\nonumber \\
 & +\mathrm{cot}(\phi_{s})\sum_{i=s}^{D}P_{i}^{2}\mathrm{cos}^{2}(\phi_{i+1})]\nonumber \\
 & =0,
\end{align}
where we have used $\sum_{i=s}^{D}P_{s}^{2}cos^{2}(\phi_{i+1}))=P_{s}^{2}$.

When $k=s$, 
\begin{align}
\frac{\partial u^{T}}{\partial\phi_{k}}\cdot\frac{\partial u^{T}}{\partial\phi_{k}} & =\sum_{i=k-1}^{D}\frac{\partial u_{i}^{T}}{\partial\phi_{k}}\frac{\partial u_{i}^{T}}{\partial\phi_{k}}\nonumber \\
 & =P_{k-1}^{2}.
\end{align}

Therefore, 
\begin{equation}
\mathbb{E}[\partial_{\phi_{k}}H\partial_{\phi_{s}}H]=\frac{r^{2-2\delta_{k1}}}{N}P_{k-1}^{2}\delta_{ks}.
\end{equation}

Thus the derivative matrices are independent centered GUE matrices
with variances $1/N$ for $\partial_{r}H$ and $r^{2}P_{k-1}^{2}/N$
with $k>1$.

We can now compute the mean QGT components in these coordinates. For
$k>1$,

\begin{equation}
\mathbb{E}[G_{\phi_{k}\phi_{k}}^{(n)}]=\frac{r^{2}P_{k-1}^{2}}{N}\sum_{m\not=n}^{N}\mathbb{E}[\frac{1}{(H_{0}+rH_{1})_{nm}^{2}}]_{H_{0}+rH_{1}}.
\end{equation}

The combined matrix $H_{0}+rH_{1}$ is again a GUE matrix, but with
variance enlarged by a factor $1+r^{2}$. Rescaling its eigenvalues
gives

\begin{equation}
\mathbb{E}[G_{\phi_{k}\phi_{k}}^{(n)}]=\frac{(N-1)r^{2}}{2(r^{2}+1)}P_{k-1}^{2},
\end{equation}
where we used the GUE identity \citep{mehta2004random,sharipov2026hilbert}

\begin{equation}
\sum_{m\not=n}^{N}\mathbb{E}[\frac{1}{(E_{n}-E_{m})^{2}}]=N(N-1).
\end{equation}

Similarly, for the radial component, 
\begin{equation}
\mathbb{E}[G_{rr}^{(n)}]=\frac{(N-1)}{2(r^{2}+1)^{2}}.
\end{equation}

Finally, the trace must be taken with the inverse metric in hyperspherical
coordinates. Since

\begin{equation}
ds^{2}=dr^{2}+\sum_{k=2}^{D}r^{2}P_{k-1}^{2}d\phi_{k}^{2},
\end{equation}
we find
\begin{align}
\mathbb{E}[S] & =\mathbb{E}[g^{ij}G_{ij}]\nonumber \\
 & =\frac{(N-1)(D+(D-1)r^{2})}{2(r^{2}+1)^{2}}.
\end{align}

\section{dominance of small resonant blocks in strongly disordered systems}

In this appendix we give a simple estimate showing why small effective
matrix dimensions naturally dominate resonant states in a strongly
disordered Anderson-type system. The purpose is not to derive the
full random-matrix ensemble from a microscopic model, but to justify
why the finite-$N$ cases studied in the main text are physically
relevant for rare resonant eigenstates in disordered systems.

Consider an Anderson-type Hamiltonian
\begin{equation}
H(\mathbf{y})=\sum_{i}\epsilon_{i}|i\rangle\langle i|+\sum_{\langle ij\rangle}J_{ij}(\mathbf{y})|i\rangle\langle j|,
\end{equation}
where the onsite energies are independent random variables. We write
$\epsilon_{i}=Wx_{i},$where $W$ measures the disorder strength and
$x_{i}$ has a bounded density $\rho(x)$, with $\rho_{\infty}\equiv\sup_{x}\rho(x)<\infty.$
The hopping amplitudes $J_{ij}$ are assumed to be short-ranged and
bounded, $|J_{ij}|\leq J_{0}.$ In the strong-disorder regime$W\gg J_{0}$,
most eigenstates are localized. Nevertheless, rare pairs or clusters
of sites can become resonant when their onsite energy mismatch is
comparable to the hopping amplitude.

To quantify this statement, define a resonant graph $\mathcal{G}_{{\rm res}}=(V,E_{{\rm res}}),$where
the vertices are lattice sites and an edge $(i,j)$ belongs to $E_{{\rm res}}$
if
\begin{equation}
|\epsilon_{i}-\epsilon_{j}|\leq A|J_{ij}|.
\end{equation}

Here $A$ is a constant of order one, specifying the resonance criterion.
Let $C(i)$ denote the connected resonant cluster containing site
$i$. If $|C(i)|\geq k$, then there exists at least one connected
tree $T$ with $\ensuremath{k}$ vertices containing $i$, such that
all $k-1$ edges of $T$ are resonant. Therefore
\begin{equation}
\{|C(i)|\geq k\}\subset\bigcup_{T\in\mathcal{T}_{i,k}}R_{T},
\end{equation}
where $\mathcal{T}_{i,k}$ is the set of connected trees with $k$
vertices containing $i$, and $R_{T}$ is the event that every edge
of $T$ is resonant.

Let $z$ be the coordination number of the lattice. A standard depth-first-search
estimate gives
\begin{equation}
|\mathcal{T}_{i,k}|\leq z^{2(k-1)}.
\end{equation}

For a fixed neighboring site $i$, the conditional probability that
the edge $(i,j)$ is resonant satisfies
\begin{equation}
\mathrm{Pr}\left(|\epsilon_{i}-\epsilon_{j}|\leq A|J_{ij}|\,\middle|\,\epsilon_{i}\right)\leq\frac{2A|J_{ij}|}{W}\rho_{\infty}\leq\frac{2AJ_{0}}{W}\rho_{\infty}.
\end{equation}

Iterating this conditional bound along a tree $T$ with $k-1$ edges
gives
\begin{equation}
\mathbb{P}(R_{T})\leq\left(\frac{2AJ_{0}\rho_{\infty}}{W}\right)^{k-1}.
\end{equation}

Using the union bound, we obtain
\begin{equation}
\mathbb{P}(|C(i)|\geq k)\leq\left(\frac{2Az^{2}J_{0}\rho_{\infty}}{W}\right)^{k-1}.
\end{equation}

Equivalently, when $W>W_{c}\equiv2Az^{2}J_{0}\rho_{\infty},$ the
probability of a $k$-site resonant cluster is exponentially suppressed:
\begin{equation}
\mathbb{P}(|C(i)|\geq k)\leq\exp[-(k-1)/\xi_{{\rm res}}],
\end{equation}
with $\xi_{{\rm res}}^{-1}=\ln\left(\frac{W}{2Az^{2}J_{0}\rho_{\infty}}\right)>0.$

This estimate shows that, in the strong-disorder regime, large resonant
clusters are rare. The dominant nontrivial resonant objects are therefore
small clusters. A single isolated localized state does not produce
nontrivial level hybridization. The first nontrivial case is a two-site
resonance, corresponding to an effective $N=2$ Hamiltonian. The next
correction is a three-site resonance, corresponding to an effective
$N=3$ Hamiltonian. Larger effective matrix dimensions are possible,
but their probability is parametrically suppressed by powers of $J_{0}/W$.
\end{document}